\documentclass[fleqn,usenatbib]{mnras}

\usepackage{newtxtext,newtxmath}
\usepackage[T1]{fontenc}
\usepackage{ae,aecompl}

\usepackage{graphicx}	% Including figure files
\usepackage{amsmath}	% Advanced maths commands
\usepackage{amssymb}	% Extra maths symbols

\newcommand{\msun}{\ensuremath{M_{\odot}}}
\newcommand{\kms}{\ensuremath{{\rm km}\,{\rm s}^{-1}}}
\newcommand{\ma}{\ensuremath{M_{\rm A}}}
\newcommand{\mb}{\ensuremath{M_{\rm B}}}
\newcommand{\mrem}{\ensuremath{M_{\rm rem}}}

\newcommand{\ai}{\ensuremath{a_{\rm AB,i}}}
\newcommand{\af}{\ensuremath{a_{\rm AB,f}}}

\newcommand{\pabi}{\ensuremath{P_{\rm AB,i}}}

\newcommand{\aaf}{\ensuremath{a_{\rm A,f}}}

\newcommand{\pai}{\ensuremath{P_{\rm A,i}}}
\newcommand{\paf}{\ensuremath{P_{\rm A,f}}}

\newcommand{\tlk}{\ensuremath{t_{\rm LK}}}
\newcommand{\tml}{\ensuremath{t_{\rm ml}}}
\newcommand{\vpn}{\ensuremath{v_{\rm PN}}}
\newcommand{\vorb}{\ensuremath{v_{\rm orb}}}
\newcommand{\vab}{\ensuremath{v_{\rm AB}}}

\title[The triple-star origin of PN Sh 2-71]{On the triple-star origin of the planetary nebula Sh~2-71}

\author[D. Jones et al.]{
David Jones,$^{1,2}$\thanks{E-mail: djones@iac.es}
Ond\v rej Pejcha,$^{3}$\thanks{E-mail: pejcha@utf.mff.cuni.cz}
and Romano L.~M. Corradi$^{4,1}$\thanks{E-mail: romano.corradi@gtc.iac.es}
\\
$^{1}$Instituto de Astrof\'isica de Canarias, E-38205 La Laguna, Tenerife, Spain\\
$^{2}$Departamento de Astrof\'isica, Universidad de La Laguna, E-38206 La Laguna, Tenerife, Spain\\
$^{3}$Institute of Theoretical Physics, Faculty of Mathematics and Physics, Charles University, 180 00 Prague, Czech Republic\\
$^{4}$GRANTECAN, Cuesta de San Jos\'e s/n, 38712 Bre\~na Baja, La Palma, Spain
}

\date{Accepted XXX. Received YYY; in original form ZZZ}

\pubyear{2019}

\begin{document}
\label{firstpage}
\pagerange{\pageref{firstpage}--\pageref{lastpage}}
\maketitle

\begin{abstract}
Recent studies have indicated that triple star systems may play a role in the formation of an appreciable number of planetary nebulae, however only one triple central star is known to date (and that system is likely too wide to have had much influence on the evolution of its component stars).  Here, we consider the possibility that Sh~2-71 was formed by a triple system which has since broken apart.  We present the discovery of two regions of emission, seemingly aligned with the proposed tertiary orbit (i.e. in line with the axis formed by the two candidate central star systems previously considered in the literature). We also perform a few simple tests of the plausibility of the triple hypothesis based on the observed properties (coordinates, radial velocities, distances and proper motions) of the stars observed close to the projected centre of the nebula, adding further support through numerical integrations of binary orbits responding to mass loss.  Although a number of open questions remain, we conclude that Sh 2-71 is currently one of the best candidates for planetary nebula formation influenced by triple-star interactions.
%the triple hypothesis cannot be ruled out and may indeed provide a reasonable explanation for some of the observed properties of Sh~2-71.
\end{abstract}


%%%%%%%%%%%%%%%%%%%%%%%%%%%%%%%%%%%%%%%%%%%%%%%%%%

%%%%%%%%%%%%%%%%% BODY OF PAPER %%%%%%%%%%%%%%%%%%

\section{Introduction}
\label{sec:intro}

Central star binarity is now the favoured hypothesis for the origins of the most axisymmetric structures found in planetary nebulae \citep[PNe;][]{jones17c}.  However, roughly 10\% of solar-type main-sequence stars are found to exist in higher order systems \citep{raghavan10}, meaning that these systems may too have an important role to play in PN formation.

In spite of the apparent support for the importance of triples in the formation of PNe, only one confirmed triple central star is known - that of NGC~246 \citep{adam14} which is, in fact, so wide that it is unlikely to have played a role in the shaping of the nebula \citep{bear17}.  Several other candidates have been found, but none has stood up to rigorous study.  The central star of SuWt~2 was frequently cited as a strong candidate triple central star, presenting with a bright binary comprising two, near-identical, A-type stars close to its projected centre \citep{bond02,exter10,jones10a}.  However, the long-term radial velocity study of \citet{jones17a} led the authors to conclude that the A-type binary was merely a field system found in chance alignment.  Similarly, \citet{boffin18} found the bright, main-sequence binary close to the projected centre of M~3-2 to also be a chance alignment. A, perhaps, promising candidate is Abell~63 (A~63), the central star of which is known to be a close-binary \citep[being the first such system to be discovered;][]{bond76}, but more recently \citet{ciardullo99} identified a nearby star (roughly 2.8\arcsec{} away from the central binary) as a possible wide tertiary companion (finding only a 1.5\% chance that the alignment is the result of a chance superposition).  The morphology of A~63, however, is rather canonical for a post-CE PN - presenting with a central bipolar/cylindrical region and higher velocity polar ejections, all sharing the same symmetry axis which in turn is perpendicular to the binary orbital plane \citep{mitchell07,hillwig16}.  Thus, the PN morphology offers no indication that the possible tertiary companion has played any role in the formation of the nebula itself.

\citet{soker16} demonstrated that perhaps the lack of known surviving triple central stars could be a result of the difficulties of surviving a common envelope (CE) phase.  In some cases, the binary system may be completely destroyed, merging with the nebular progenitor, or could merge with one another to leave a single companion \citep{hillel17}.  Furthermore, due to tidal forces and mass-loss as the nebular progenitor ascends the AGB, the stability of the system may be reduced even before reaching a CE phase, perhaps leading to the ejection of one component of the binary system, leading to a wide variety of possible central star configurations \citep[as discussed in detail in section 3.2.1 of][]{soker16}.

Sh~2-71 ($\alpha=19^h02^m00.29^s$,$\delta=+02^\circ09' 10.97''$, PN~G035.9$-$01.1) was discovered by  \citet{minkowski46} and classified as a ``diffuse and peculiar'' nebulosity.  The object was then later included in the catalogue of H\textsc{ii} regions by \citet{sharpless59} with the caveat that it may be a PN -- a classification which has since been made more definitive by later spectroscopic studies \citep{chopinet76,bohigas01}.  The detailed spectroscopic study of \citet{bohigas01} revealed the PN to be of Peimbert Type \textsc{i},  showing strong signs of shock excitation and significant density variations across the nebula. Furthermore, they reported the requirement for an extremely hot central star ($T\sim 200\,000$\,K) derived via the crossover method \citep{kaler89a}.  The energy-balance method of \citet{preite89} returns a lower temperature of $\sim 130\,000$ K, consistent with estimate of \citet{feibelman99} based on the comparison of archival IUE spectra with similar data from other hot central stars\footnote{It is important to note that the IUE aperture covers a 10\arcsec{}$\times$23\arcsec{} region which includes both binary A and star B.}.  

\begin{figure}
\centering
	\includegraphics[angle=0,height=7.5cm, trim=100 0 100 0, clip]{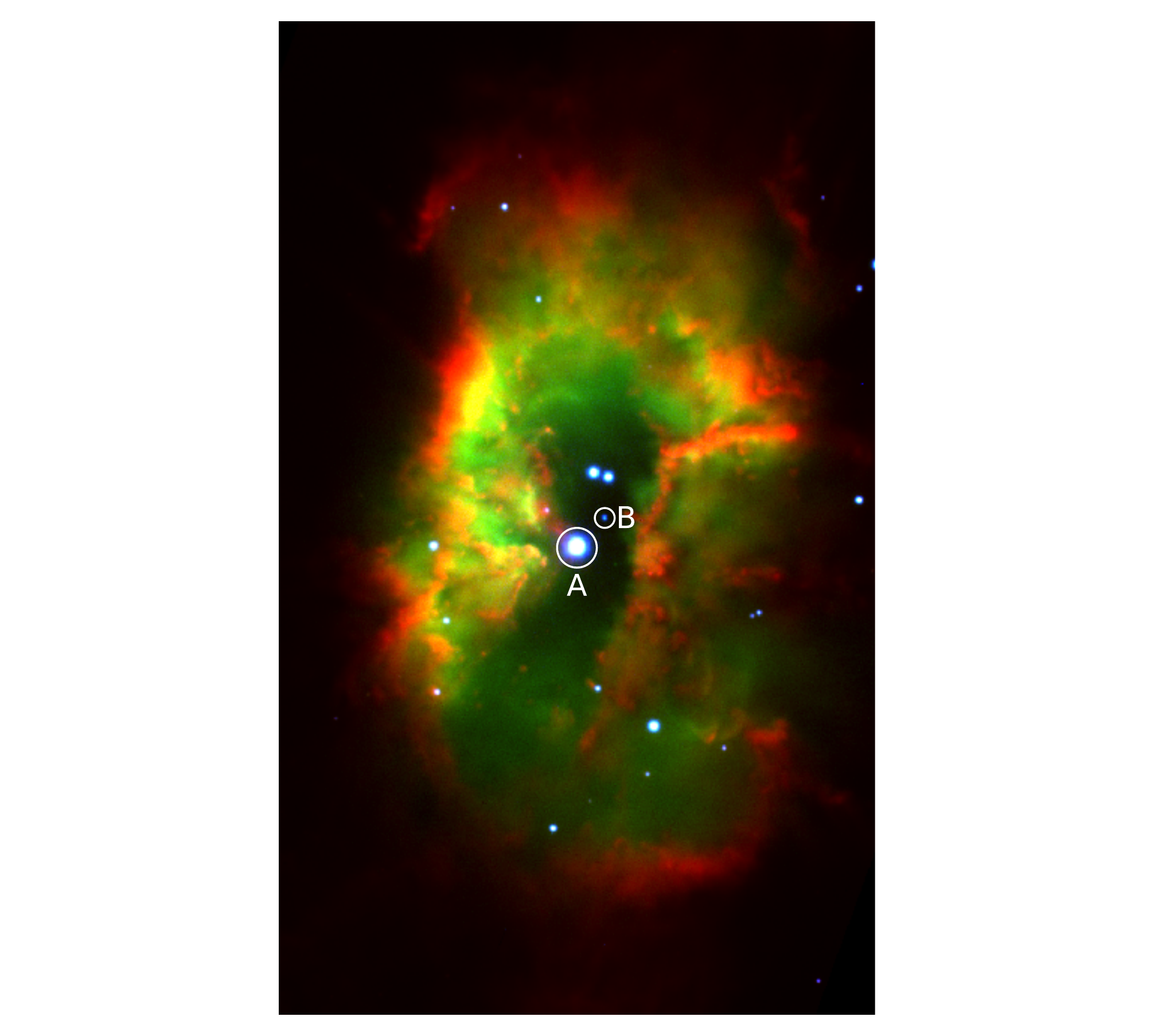}
    \caption{Colour-composite image of the central region of Sh~2-71 produced from Gemini-GMOS images taken from the Gemini archive (H$\alpha$ is red, [O\textsc{iii}] is green, He~\textsc{ii} is blue - each image was of 300s exposure time).  North is up, East is left.  The image measures roughly 1.8\arcmin{}$\times$3\arcmin{}. The candidate central stars, the bright binary star \citep[labelled A;][]{mocnik15} and the fainter blue star to its North \citep[labelled B;][]{frew07}, are both circled.
    }
    \label{fig:gemini_im}
\end{figure}

All of the aforementioned studies confirm the need for a hot central star, however the bright star observed close to the nebular centre (labelled A in figure \ref{fig:gemini_im}) was found by \citet{kohoutek79} to present with colours consistent with a B8V classification.  \citet{kohoutek79} also showed the star to be variable, concluding that it was likely in a close binary system with the hot nebular progenitor.  Since then, there has been much debate around the identity of the true central star, with \citet{frew07} suggesting that the faint, blue star to the North-West of A (labelled B in figure \ref{fig:gemini_im}) might be a better candidate.  The detailed analysis of A by \citet{mocnik15} concluded it to be a rather exotic system consisting of a Be binary with a misaligned, precessing disc.  They surmise that the companion could be a low-mass subdwarf which would then be the nebular progenitor, however they find no definitive evidence for such a hot component in the system.

Based on the nebular morphology, \citet{bear17} classify Sh~2-71 as likely to have originated from a triple system, claiming that the pronounced lack of axial and/or mirror-symmetry is characteristic of such interactions.  Furthermore, the hydrodynamical simulations of \citet{akashi17} show that jets launched from a binary system in an inclined orbit with a tertiary AGB companion result in PN morphologies and density variations remarkably similar to those observed in Sh~2-71.  This leads us to consider here the possibility that both binary A and star B form or once formed a triple system, the interacting evolution of which led to the formation of Sh~2-71.  Section \ref{sec:emission} presents the discovery of two extended regions of emission several arcminutes from the central PN shell - the formation of which may be related to the interacting history between binary A and star B.  In section \ref{sec:triple}, we consider the possible history of the A-B system and assess the plausibility of such a configuration resulting in the currently observed positions and proper motions, while in section \ref{sec:discussion} we conclude with a discussion and outline of possible future analyses.

\section{Extended emission regions}
\label{sec:emission}

Sh~2-71 was observed on 2011 April 25 using the Wide Field Camera (WFC) instrument mounted on the the 2.5-m Isaac Newton Telescope (INT) at the Observatorio del Roque de Los Muchachos on the Spanish island of La Palma.  Seven 400s exposures were acquired through the H$\alpha$+[N\textsc{ii}] filter (ING filter ID\#197, central wavelength of 6568\AA{}, FWHM of 95\AA{}) with binning 1$\times$1 for a pixel scale of 0.33\arcsec{}$\times$0.33\arcsec{}, where the four mosaiced chips of the WFC cover a field of view of approximately 34\arcmin{}$\times$34\arcmin{}.  The resulting images were debiased, flat-fielded and stacked using standard \textsc{iraf} routines \citep{iraf}.

\begin{figure*}
	\includegraphics[width=\textwidth]{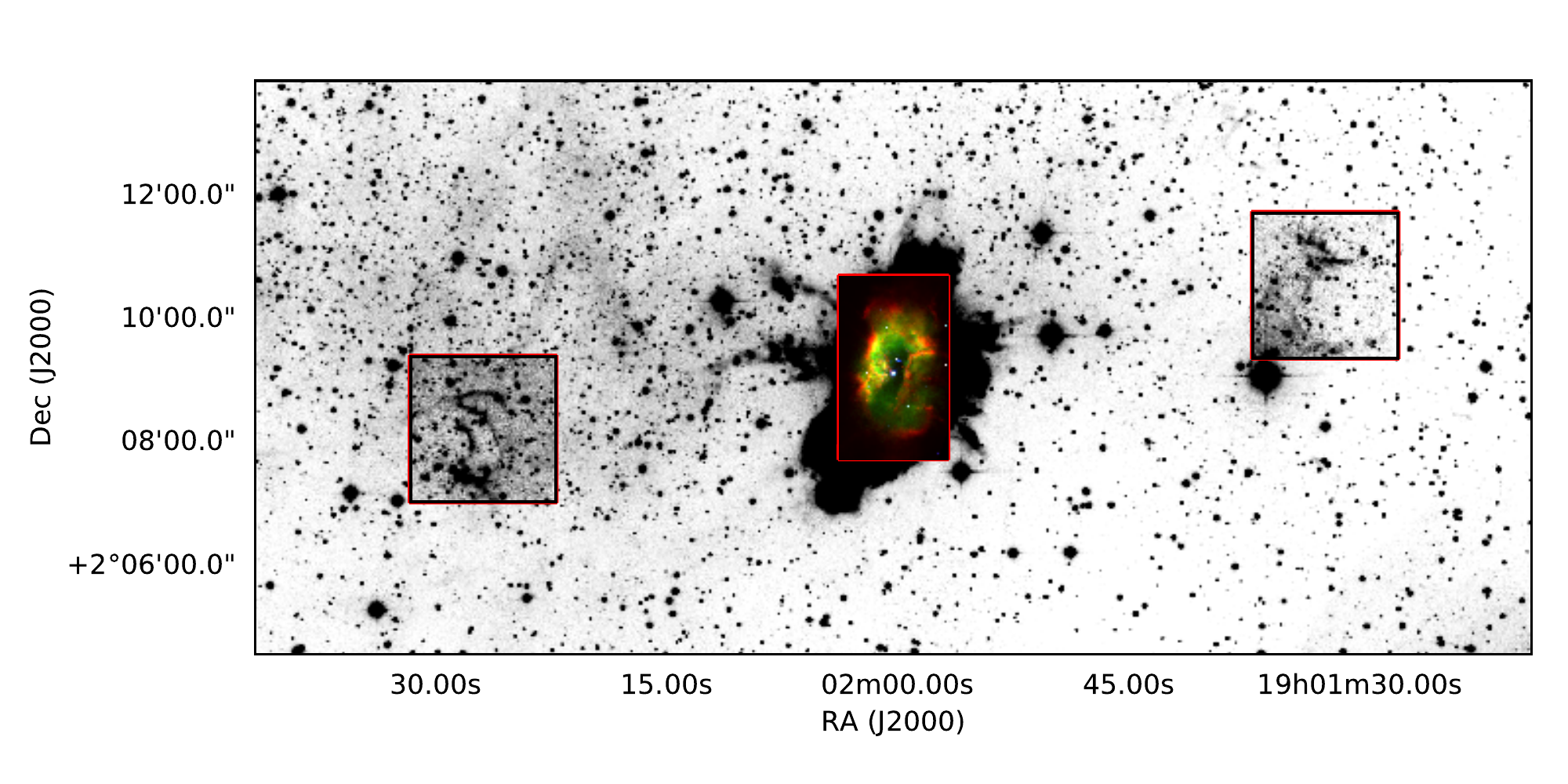}
    \caption{INT-WFC image of Sh~2-71 taken in the light of H$\alpha$+[N\textsc{ii}] highlighting the presence of the extended regions of emission to the East and West of the central nebula (the colour composite of figure \ref{fig:gemini_im} is overlaid to demonstrate the the full extent of the central structures).  To highlight the filamentary structures present in the newly discovered emission regions against the diffuse background emission, they are shown as cut-outs with different display stretches.
    }
    \label{fig:wfc_im}
\end{figure*}

The deep and wide-field nature of the observations reveal, for the first time, the presence of extended knots of emission several arcminutes away from the central nebula as highlighted in figure \ref{fig:wfc_im}.  The bandpass of the filter employed covers H$\alpha$ as well as both [N\textsc{ii}]~6548\AA{} and 6583\AA{} lines, as such it is not clear as to whether the knots are H$\alpha$-bright, [N\textsc{ii}]-bright or both. The features present as faint filamentary structures superimposed on the diffuse background emission associated with the nearby H\textsc{ii} region located to the East \citep[KC97c~G036.3$-$01.7;][]{KC97}.  As such, in Figure \ref{fig:wfc_im} the two emission regions are shown as cut-outs with different display stretches in order to fully outline their structures against the varying diffuse background (which shows an appreciable gradient across the image).

The emission features are found in roughly the East-West direction, almost perpendicular to the apparent symmetry axis of the central nebular region which extends North-South. Intriguingly, the emission regions lie relatively close to line connecting binary A and star B, with the position angle between the two knots of emission being roughly 100$^\circ$ while the position angle connecting star B to binary A is 136$^\circ$.  Given the location of the newly discovered features (approximately equidistant from and symmetrically placed around the centre of Sh~2-71) as well as their apparent structure (filamentary c.f.\ the diffuse background), we conclude that they are indeed related to Sh~2-71 and perhaps a consequence of the evolution of the central star(s).  We will elaborate further on possible formation scenarios for these features in Sec.~\ref{sec:ext_origins}.

\section{The triple hypothesis}
\label{sec:triple}

\begin{figure*}
    \centering
    \includegraphics[width=0.7\textwidth]{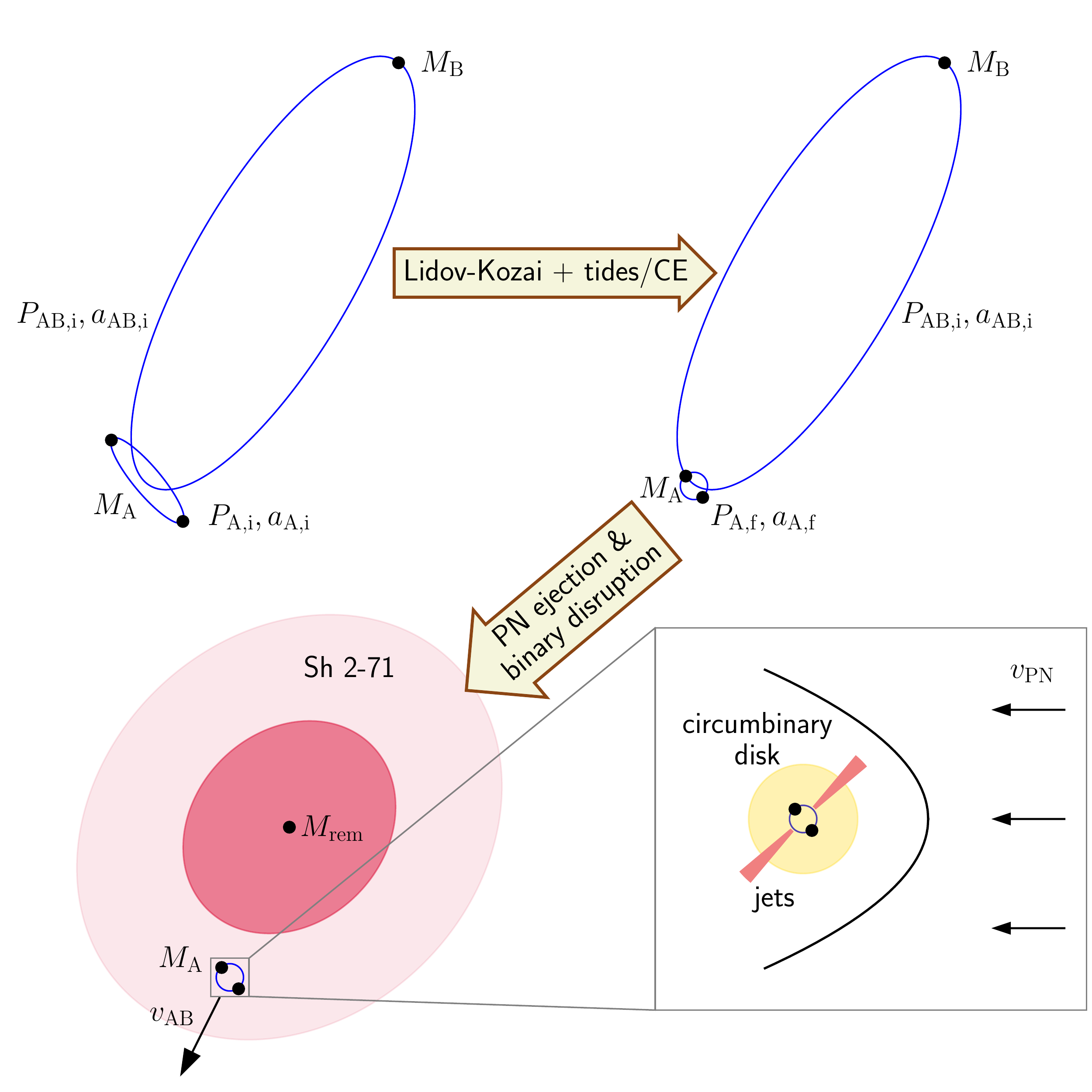}
    \caption{Summary of our fiducial triple model for Sh 2-71 and stars A and B. We start with an hierarchical triple system, where the inner binary A with period $\pai$ is orbited by companion B  with orbital period $\pabi \gg \pai$. Star B causes Lidov-Kozai oscillations in binary A, which are accompanied by tides, mass transfer, or common envelope evolution. This leads to formation of a short-period binary, $\paf \ll \pai$, with peculiar photometric variability currently observed in A. Star B is initially more massive than A and ejects its envelope seen today as the PN Sh~2-71. The mass ejection disrupts the orbit of A and B and both objects fly away with velocity $\vab$. Mass ejected from B passes through A, where part of it is captured, which might lead to formation of circumbinary disk, jets, or other fast ejections \citep{soker04}.}
    \label{fig:diagram}
\end{figure*}

In Sh~2-71, a binary star with a very peculiar variability pattern but insufficient temperature to ionize the nebula  (star A) is projected close to another hot star. The nebula Sh~2-71 itself exhibits an irregular morphology, which has been attributed to interactions in a close triple star system.
One may thus consider the hypothesis that at some time the binary system A was much closer, and indeed bound, to the faint, (likely) nebular progenitor, star B.  For this to be the case, binary A and star B would have to be at approximately the same distance.  \textit{Gaia} parallaxes put binary A at a distance of 1.62$^{+0.10}_{-0.09}$ kpc \citep{gaia_satellite,bailerjones18}, but unfortunately the faintness of star B prevented a meaningful parallax measurement.  As a proxy, one can use the distance to the PN itself.  While PN distances are notoriously difficult to derive, the H$\alpha$ surface brightness-radius relationship of \citet{frew16} places Sh~2-71 at a distance of $1.52 \pm0.54$\,kpc -- consistent with that of binary A. Stars A and B are separated by $7.4\arcsec{}$, which corresponds to a minimum physical separation of $1.2\times 10^4$\,AU at this distance.

Binary A was found by \citet{mocnik15} to comprise a 2.6 M$_\odot$ B8V star and a low-mass companion. \citet{mocnik15} did not explicitly determine the orbital period $\paf$, but estimated that it is few days. We label the semi-major axis corresponding to $\paf$ as $\aaf$.  Furthermore, \citet{mocnik15} concluded that the peculiar 68-day period photometric variability, analyzed in detail by \citet{mikulasek05,mikulasek07}, is the result of obscuration by a precessing disc on a longer period than $\paf$.  It is thus plausible that formation of such an unusual configuration requires strong mass transfer or a CE event within the binary \citep[e.g.][]{han02,han03,justham11}. Star B could have played a significant role in the formation of A either through direct mass transfer \citep{soker04} or by forcing the components of A to strongly interact by the action of Lidov-Kozai cycles \citep[e.g.][]{fabrycky07,thompson11,perets12,shappee13,pejcha13,michaely14,naoz14,naoz16}. 

In our fiducial scenario, summarized in Figure~\ref{fig:diagram}, binary A and star B were initially much closer and in a bound orbit. The original orbital period of binary A was much greater than that currently observed, $\pai \gg \paf$. The action of Lidov-Kozai cycles induced by star B would shrink the orbit of A to the current value, possibly inducing mass transfer or a CE thus explaining the peculiar properties of A. Eventually, star B ascended the AGB and lost a significant fraction of its mass, which caused star A to move farther away or end up on an unbound trajectory. We test this hypothesis with simple analytic estimates (Sec.~\ref{sec:analytic}) and numerical calculations of binary orbit breakup (Sec.~\ref{sec:rebound}). We  perform more consistency checks with observations (Sec.~\ref{sec:checks}) and discuss variations on our fiducial triple hypothesis (Sec.~\ref{sec:variations}).

%In this scenario, binary A and star B would have initially been much closer, but as star B ascended the AGB (losing a significant fraction of its mass) binary A became unbound.  

%\citet{perets12} showed that evolved triple systems experience significant instabilities (due to mass-loss and/or mass transfer) which almost invariably lead either to collisions or ejections, like the one considered here. 

\subsection{Analytic estimates}
\label{sec:analytic}

If the total mass of binary A is $\ma \sim 3 \msun$ \citep[consistent with the estimates of][]{mocnik15}, then the central star B should have initially been rather massive, $\mb \ge \ma$, to have contemporaneously evolved to its current state\footnote{Technically, star B need only be more massive than the most massive component of A. However, since A likely experienced a rather complex mass transfer history, we consider the conservative case that the evolution of A roughly depends on its total mass.}.  This high mass would imply that star B has left a roughly $\mrem \approx 1\,\msun$ remnant. Isotropic mass loss from one component of a binary system will cause the orbit to widen or even break apart. The conditions separating these two outcomes were summarized by \citet{michaely19}. 

Let us first consider the case of binary breakup. This happens, for example, when a binary on a circular orbit instantly loses more than half of its total mass. Consequently, we assume that $\mb \sim 5\,\msun$ noting that eccentricity and slow mass loss will allow a range of $\mb$, as we show in Section~\ref{sec:rebound}. After breakup, the two components fly away with velocities similar to their instantaneous orbital velocity,  $\vab \sim \vorb$. From this velocity, we can put a constraint on the original orbital semi-major axis of A and B, $\ai$. The observational constraints on the relative velocity of A and B are uncertain and are discussed in more detail in Section~\ref{sec:checks}. For the purposes of analytic estimates, we assume that the relative velocity is $\vab \approx 4\,\kms$ (see Sec.~\ref{sec:checks} for the derivation of this estimate). The semimajor axis $\ai$ corresponding to this velocity, assuming circular orbit, was thus
\begin{equation}
\ai \sim \frac{G(M_A+M_B)}{\vab^2} \sim 440\,{\rm AU} \left( \frac{8\,M_\odot}{M_A+M_B}\right) \left(\frac{4\,{\rm km\,s^{-1}}}{\vab}\right)^{-2},
\label{eq:a}
\end{equation}
and the orbital period was
\begin{equation}
\pabi = \frac{2\pi \ai}{\vorb} \sim \frac{2\pi \ai}{\vab} \sim 3300\,{\rm yr} \left( \frac{8\,M_\odot}{M_A+M_B}\right) \left(\frac{4\,{\rm km\,s^{-1}}}{v_{\rm orb}}\right)^{-3}.
\label{eq:p}
\end{equation}
The constraint on $\pabi$ is very sensitive to $\vorb$, which in turn depends on uncertain spatial kinematics of binary A, star B, and the PN.

Returning to the possibility that a Lidov-Kozai interaction between binary A and star B could have resulted in the unusual properties of binary A, the timescale connected with such interactions is 
\begin{eqnarray}
\tlk &=& \frac{8}{15\pi} \frac{M_A + M_B}{M_B}\frac{\pabi^2}{\pai} (1-e_{\rm AB}^2)^{3/2} \sim \nonumber \\
& \sim & 11\,{\rm Myr}\, (1-e_{\rm AB}^2)^{3/2} \left(\frac{\pai}{100\,{\rm days}}  \right)^{-1},
\label{eq:kozai}
\end{eqnarray} 
where $e_{\rm AB}$ is the eccentricity of the outer orbit \citep{naoz16}. In this estimate, we assumed a rather arbitrary value of $\pai = 100$ days to allow for easy rescaling. If Lidov-Kozai cycles operate efficiently and generate high-eccentricity periastron passages, $\pai$ decreases due to dissipative processes such as tides or mass transfer. Ultimately, the orbital period stabilizes near its current value $\paf$. Our estimate of $\tlk$ is still considerably shorter than the lifetime of a $5\,\msun$ star, at around 100 Myr \citep[Eq. 2.4]{eggleton06}. 
%Furthermore, given that a system generally needs several tens of $t_K$ for Kozai-Lidov interactions to significantly alter the system's properties, the hypothesis would seem to be plausible based on the considered timescales.

The mass loss from B could have been sufficiently slow or the total mass lost too small to break up the binary. In this case, stars A and B are still bound on a wide orbit. Without knowing how much mass was lost, we have no handle on $\ai$ and hence no constraint on $\tlk$. However, a bound orbit gives a very specific prediction for the mutual velocity $\vab$ of two stars with separation $r$
\begin{equation}
    \vab^2 = G(\ma + \mrem) \left(\frac{2}{r} - \frac{1}{\af} \right).
    \label{eq:vorb}
\end{equation}
In our case, $\mrem = 1\,\msun$, $r \gtrsim 1.2 \times 10^4$\,AU from the projected position on the sky, and $2\af > r$ from the properties of an elliptical orbit. If the mutual velocity of A and B is found significantly in excess of the highest possible velocity given $r$, $\sqrt{2G(\mb+\mrem)/r} \approx 0.8\,\kms$, then A and B cannot be bound.

\subsection{Numerical integrations}
\label{sec:rebound}

\begin{figure*}
    \centering
    \includegraphics[width=0.8\textwidth]{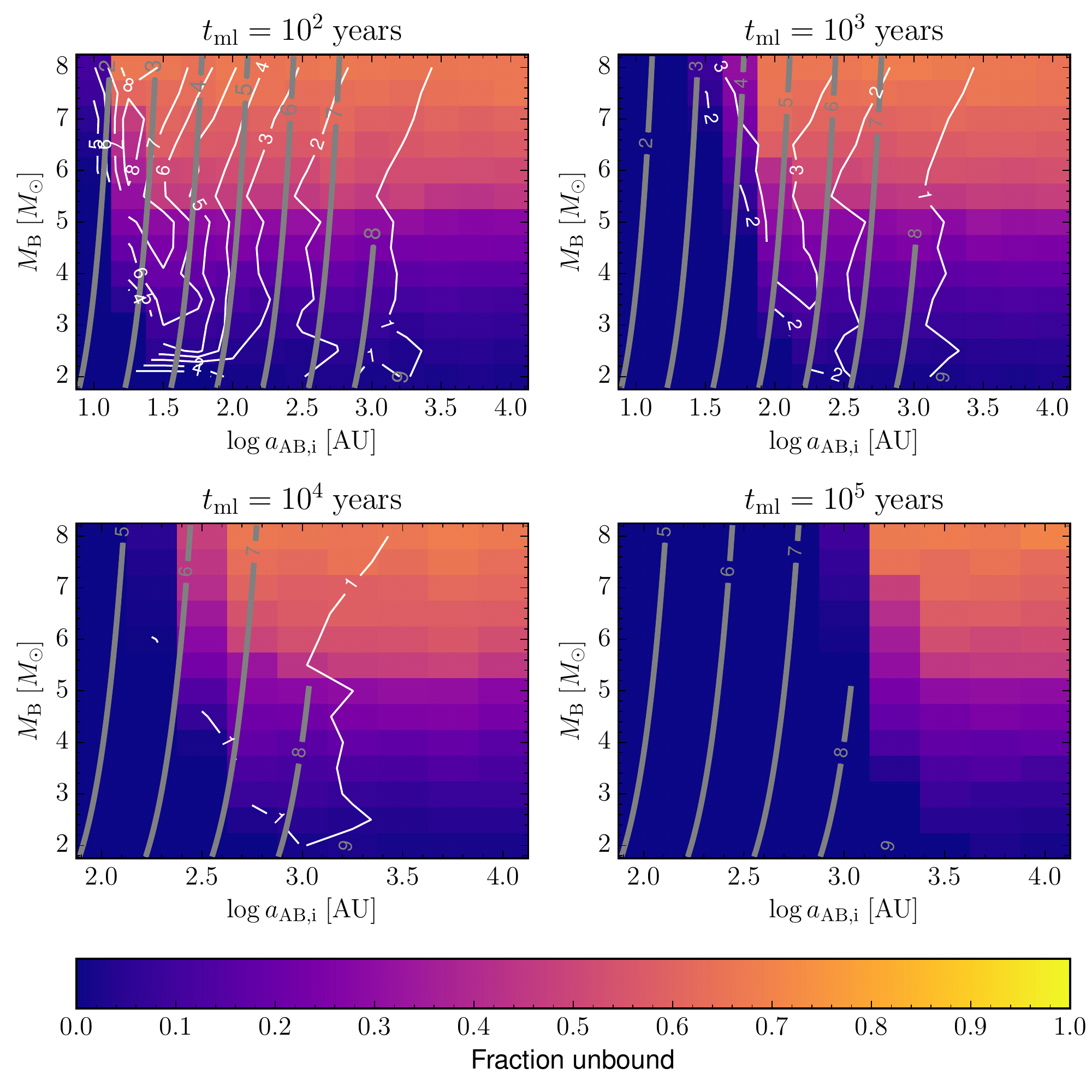}
    \caption{Fraction of disrupted binaries as a function of the initial mass of the mass-losing component $\mb$ and initial semi-major axis $\ai$. The four panels are for different mass-loss timescales $\tml$, over which star B experiences a constant mass-loss rate. White contours indicate median mutual velocity of the two disrupted components $\vab$, labeled in \kms. Gray contours show the Lidov-Kozai timescale $\tlk$ (Eq.~[\ref{eq:kozai}]) for $\pai = 100$\,days, but only for initial conditions where $\tlk$ is shorter than the main-sequence lifetime of star B (thus these contours delimit the parameter space in which Lidov-Kozai cycles could have an appreciable effect on the binary evolution). Contours are labeled in $\log_{10} \tlk$ in years.}
    \label{fig:binary}
\end{figure*}

\begin{figure*}
    \centering
    \includegraphics[width=0.8\textwidth]{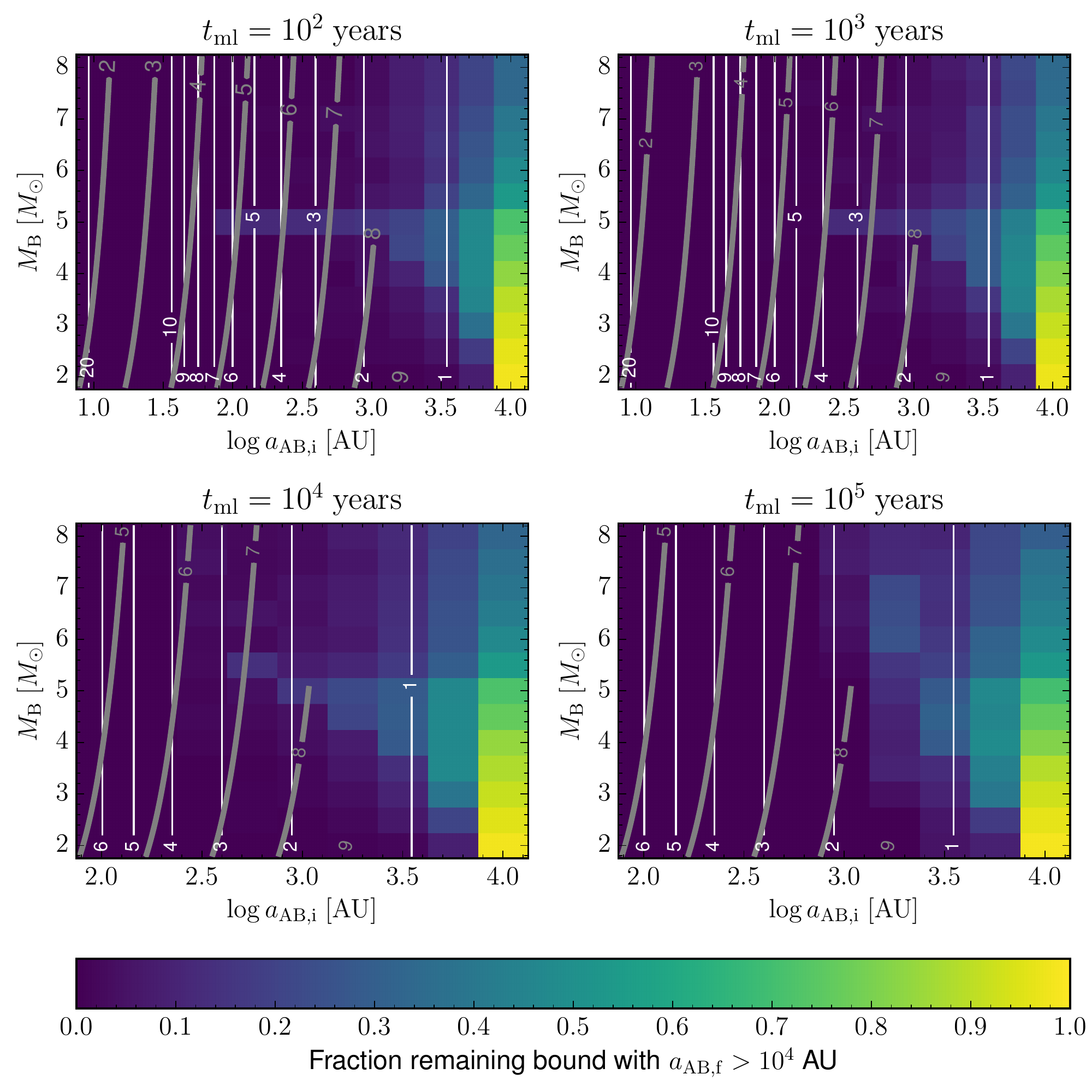}
    \caption{Fraction of initial conditions that remain bound and end up with $\af \ge 10^4$\,AU as a function of initial $\mb$, $\ai$, and for four different $\tml$. White contours show the typical orbital velocity of the surviving binary, $\sqrt{G(\ma+\mb)/\af}$, where $\ma+\mb = 4\,\msun$. Contours are labeled in \kms. Gray contours show $\tlk$ similarly to Figure~\ref{fig:binary}.}
    \label{fig:binary_bound}
\end{figure*}

Evolution of hierarchical triples (or higher multiplicity systems) with tidal effects, stellar evolution, mass transfer, and  CE evolution is challenging and has only been attempted in a handful of cases \citep{hamers13,hamers18,hamers19,lu19}. If mass ejection occurs, the commonly employed double averaging of the orbits breaks down and the system should be studied with direct integration. Focusing on this dynamical phase, \citet{michaely19} studied the orbital properties of bound companions to post-CE binaries and constrained the duration of CE to $t_{\rm ml} \sim 10^3$ to $10^5$\,years. Following their ideas, we can quantify the constraints illustrated in Section~\ref{sec:analytic} and place limits on the mass ejection timescale $t_{\rm ml}$ from star B.

Inspired by \citet{michaely19}, we employed the software package {\sc Rebound} \citep{rein12} to calculate evolution of two orbiting point masses. The binarity of star A was neglected as are further degrees of freedom and complications (spins, tides, general relativity, stellar evolution, outside perturbations, etc.). Mass $\ma = 3\,\msun$ is held fixed and mass $\mb$ decreases linearly over time $\tml \in \{10^2, 10^3, 10^4, 10^5\}$\,years from its initial value to $\mrem = 1\,\msun$. This is achieved with function {\tt post\_timestep\_modifications}. The binary is moved to its center of mass after every timestep using function {\tt reb\_move\_to\_com}. The initial values of $\mb$ were varied on a regular grid from $2$ to $8\,\msun$. Initial semi-major axes $\ai$ were varied on a log-uniform grid from $10^2$ to $10^4$\,AU. Initial eccentricities were set to 9 values uniformly spaced between $0$ and $0.9$. For each set of $\mb$, $\ai$, and eccentricity, we randomly drew $100$ values of mean anomaly from a uniform distribution and integrated using {\sc WHFast} \citep{rein15}. After rescaling of $\mb$, our results are generally applicable to triple systems of similar hierarchy. 

In Figure~\ref{fig:binary}, we present the fraction of disrupted binaries as a function of initial $\mb$, $\ai$, and $\tml$ when marginalized over initial eccentricity and true anomaly. Binaries disrupt only if $\tml$ is approximately shorter than the initial orbital period, as expected. Even binaries which lose less than half of their total mass can be disrupted for certain eccentricities, but the probability is lower. For each initial $\mb$ and $\ai$ we also show the median mutual velocity of the binary after breakup (white contours). Binaries with larger $\ai$ lead to slower disruption, but mutual velocities also decrease close to bound/unbound boundary. There is order of unity scatter in the mutual velocities after disruption. We also show contours of $\tlk$ (Eq.~[\ref{eq:kozai}]), but only when it is shorter than the main sequence lifetime of star B \citep[Eq.~2.4 of][]{eggleton06}.

In Figure~\ref{fig:binary_bound}, we show the fraction of initial conditions, which remain bound and end up with semi-major axis $\af \ge 10^4$\,AU (the current minimum separation of A and B). We see that this fraction is rather insensitive, being generally very low for our initial conditions and increasing only slightly for higher $t_{\rm ml}$ as the mass loss becomes more adiabatic. Our results suggest very small probability for $\ai \lesssim 10^3$\,AU, although these binaries remain mostly bound as seen in Figure~\ref{fig:binary}. This is because adiabatic mass loss changes $\af/\ai$ by a constant factor and these $\ai$ are too small to end up with $\af \ge 10^4$\,AU. In other words, the binary loses too little mass. Similarly to Figure~\ref{fig:binary}, we also overplot the Lidov-Kozai timescale when shorter than the main-sequence lifetime (gray contours) and the typical orbital velocity after disruption (white contours, Eq.~[\ref{eq:vorb}]).

\subsection{Observational constraints}
\label{sec:checks}

Figures~\ref{fig:binary} and \ref{fig:binary_bound} suggest that the past evolution of A and B can be constrained from their precise relative velocity. The proper motion difference of A and B from \textit{Gaia} DR2 is $2.3 \pm1.4$ mas yr$^{-1}$ \citep{gaia_satellite}, which at the distance of A (and likely also Sh 2-71) corresponds to tangential velocity of $18 \pm 11\,\kms$. The large uncertainties on the proper motion values of the faint star B have a detrimental impact on derivation of both the direction\footnote{Indeed, the direction of proper motion difference is found not to be aligned with the axis between A and B, but with particularly large uncertainty.} and magnitude of proper motion difference, limiting the usefulness of these estimate for our purposes. 

There is an alternative way of estimating the relative velocity of A and B. If the disruption of the original binary occurred at the same time as the Sh~2-71 was ejected, the relative position of A with respect to the bright nebular shell gives us an approximation of the ratio of projected velocities.  Binary A is roughly quarter-way between star B and the nearest ``wall'' of the nebula, implying that the orbital velocity at break-up could be roughly one quarter of the PN expansion velocity. The expansion velocity of the PN was found by \citet{sabbadin84} to be $\vpn \approx  16\,\kms$, implying a break-up velocity of $\vorb \sim 4$\,\kms. Finally, the difference in systemic velocities between binary A and the PN was found by \citet{mocnik15} to be of order $2\,\kms$. 

All of this information together suggests that the relative velocity of A and B is likely few $\kms$. As an illustration of what can be inferred from the relative velocity, we now assume that it was measured to be $4\,\kms$, as suggested by the relative positions of A, B and the PN. From Figure~\ref{fig:binary_bound} and Equation~(\ref{eq:vorb}), we see that a bound orbit for A and B is very unlikely. Instead, A and B have to be on an hyperbolic trajectory. Figure~\ref{fig:binary} then illustrates that the timescale for mass-loss $\tml$ had to be shorter than about $10^4$\,years. For longer $\tml$, only binaries with wide $\ai$ and slow $\vorb$ are disrupted, which result in too small relative velocities $\vab$. Finally, unless $\tml \lesssim 10^2$\,years \citep[inconsistent with typical PN formation timescales, e.g;][]{szyszka11}, a relatively high $\mb$ would be required to explain the relative velocity.

%For example, if A and B are currently unbound (Fig.~\ref{fig:binary}), $\tlk$ together with stellar lifetime determines a range of $\ai$their mutual velocity must be $\gtrsim 2\,\kms$.

%At a distance of 1.62 kpc, a velocity of 4 km~s$^{-1}$ would correspond to a proper motion of approximately $1.4$ mas yr$^{-1}$ - marginally consistent with the difference in proper motions between binary A and star B at , but not with the direction of proper motion difference which is not aligned with the axis between A and B.   Similarly, t in keeping with the relatively low velocities considered for the calculations above and in no way rules out an association (either between the PN and A or A and B).

As previously highlighted, for binary A and star B to be coeval and to accommodate the kinematical constraints on A and B, the initial mass of star B would have had to be rather massive. Sh~2-71 was classified by \citet{bohigas01} as a Peimbert type \textsc{i} nebula - a classification generally associated with massive progenitors \citep{phillips05}.

%If this were the case, then one would expect the PN to be of the Peimbert type \textsc{i} class \citep[just as found by][]{bohigas01}.

A further consistency check can be performed using the observed properties of star B.  Firstly, one can ask: are the observed colours of the star \citep[for example, those provided in the second data release of the VPHAS+ survey;][as shown in table \ref{tab:vphas}]{drew14} consistent with a post-AGB star, with a post-AGB age roughly comparable to that of the PN at $\sim10$~kyr, while originating from a sufficiently high-mass progenitor?  The $5\,\msun$ track of \citet{vassiliadis94} reaches $\log T_\mathrm{eff}$=5.2, consistent with the temperature of the ionising source required to reproduce the observed nebular spectrum  \citep[][]{bohigas01,preite89}, at approximately 10 kyr.  A blackbody of that temperature, accounting for an extinction of E(B-V)=0.64 \citep{frew16} and assuming the reddening law of \citet{cardelli89}, would have colours $u-g=-0.92$ (c.f.\ $0.97\pm0.09$ in VPHAS+), $g-r=0.34$ (c.f.\ $0.4\pm0.09$) and $r-i=0.21$ (c.f.\ $0.26\pm0.12$) - perfectly consistent with those values observed (see Table \ref{tab:vphas}).  Furthermore, scaling for the model luminosity of $\log \mathrm{L}\approx2.0$ at that point on the track and placing the system at the \textit{Gaia} distance of binary A, the predicted observed apparent magnitudes \citep[calculated using synphot;][]{synphot} shown in Table~\ref{tab:vphas} are strikingly similar to those observed especially considering the various uncertainties involved (distance, reddening, evolutionary track, blackbody assumption).

\begin{table}
	\centering
	\caption{VPHAS+ photometry of star B \citep{drew14}, alongside synphot model magnitudes for a post-AGB star \citep[of initial mass 5~M$_\odot$;][]{vassiliadis94} at the distance of binary A. }
	\label{tab:vphas}
	\begin{tabular}{lrlc} % four columns, alignment for each
		\hline
		Band & \multicolumn{2}{c}{VPHAS+} & Synphot\\
%		 & \multicolumn{2}{c}{(Vega magnitude)} & \\

		\hline
		u' & 19.23 & $\pm$0.05 & 19.27\\
		g' & 20.20 & $\pm$0.04 & 20.19\\
		r' & 19.80 & $\pm$0.05 & 19.85\\
		i' & 19.54 & $\pm$0.07 & 19.64\\
		\hline
	\end{tabular}
\end{table}

\subsection{The origins of the extended emission regions}
\label{sec:ext_origins}

Within the context of our fiducial model, it is interesting to consider the possible origins of the newly-discovered extended emission regions described in Section~\ref{sec:emission}. As the PN originating from B expands, binary A accretes material within its Bondi-Hoyle-Lyttleton radius $r_{\rm BHL} = 2GM/\vpn^2 \sim 21\,{\rm AU}\, (\vpn/16\,\kms)^{-2}$. The fraction of material accreted on A is about $r_{\rm BHL}^2/a^2 \sim 0.04\, (a/100\,{\rm AU})^{-2}$, where $a$ is the instantaneous separation of A and B. As the PN expands, $a$ increases as the binary loses mass and the PN velocity at A either stays constant or decreases depending on the PN velocity profile. The PN material captured by A will be accreted onto the binary likely through a circumbinary disk and smaller disks around individual components \citep{soker04}. Spiral modes excited in the circumbinary disk can further remove angular momentum from the orbit \citep[e.g.][]{artymowicz96,munoz19}, although this effect is likely much smaller than the previous evolution driven by Lidov-Kozai. It is tempting to speculate that the putative obscuring disk responsible for the peculiar 68-day photometric variability in A \citep{mocnik15} is a remnant of this phase, although the plausibility of this speculation depends on the disk mass and lifetime, which remain uncertain. Nonetheless, we expect that a fraction of the circumbinary material will be mixed and shocked due to binary motions and will be accelerated to leave the system with roughly the binary orbital velocity, $v_{\rm orb, A} = \sqrt{G\ma/\aaf} \sim 180\,\kms$, where $\aaf$ is the current semi-major axis of the binary in A, which we assume to have orbital period $\paf \approx 5$\,days. This is about ten times higher than $\vpn$ and the binary-accelerated material would thus be located at proportionally larger distances from the PN center. 

Figure~\ref{fig:wfc_im} shows that the extended emission is located roughly five times further out than the PN edge - vaguely consistent with the hypothesis considered providing one allows for deceleration due to interaction with surrounding interstellar medium. We note that different mass ejection mechanisms from binary stars will lead to different ejection velocities and correspondingly different positions of the putative ejecta. We expect the velocities to range from escape velocity from a stellar surface \citep[or higher if a magnetohydrodynamic jet operates;][]{huarte-espinosa12} to the escape velocity from an outer boundary of a circumbinary disk, which is lower than the binary orbital velocity. Nonetheless, $v_{\rm orb,A}$ provides a rough scaling sufficient for our order-of-magnitude estimates. A more detailed analysis would be warranted once we know with greater certainty the origin of the extended emission and the the orbital properties of star A.

The extended emission could, alternatively, be associated with binary evolution in A. One may speculate that they could be the product of non-conservative mass transfer in A, perhaps driven by Lidov-Kozai interactions with B, or dense wind from B focused into the orbital plane with A (perhaps during periastron passage, a dynamical instability, or a thermal pulse event on the AGB). However, this would require significant mass loss in A happening almost concurrently with PN ejection in B due to short visibility times associated with both the PN and the extended emission regions compared to stellar lifetimes.

\subsection{Variations on the triple hypothesis}
\label{sec:variations}

We now discuss several variations of our fiducial scenario as presented in Figure~\ref{fig:diagram}. 

\subsubsection{Inefficient Lidov-Kozai}
Firstly, binary A could have evolved to its current state independently of B. In particular, the Lidov-Kozai cycles might have been inefficient, for example due to unfavorable relative inclination of the orbits. Binary A would have evolved essentially in isolation and any strong binary evolution processes must have happened a sufficiently long time ago, because the components of A are too cool to ionise Sh~2-71 (based on the various estimates detailed in Sec.~\ref{sec:intro}).  Removing the Lidov-Kozai constraint does not significantly affect our conclusions, except that we would require $\ma \gtrsim \mb$ to have A evolve before B. This is not impossible, even though the probability of disrupting the binary by PN ejection from B is somewhat lower as can be seen from Figure~\ref{fig:binary}. The Lidov-Kozai constraint would become important for $\vab \lesssim 2\,\kms$.

\subsubsection{Triple CE evolution}

It is possible that Sh~2-71 experienced a true triple CE ejection, where all three stars strongly interacted, ejected B's envelope, and formed the peculiar binary A. We cannot exclude or straightforwardly constrain this scenario, because it has not been sufficiently theoretically explored. Although we do not require triple CE ejection to explain Sh~2-71, the odds might change in the future if more observational data cannot be accommodated within our model or its modifications. Furthermore, Sh~2-71  offers us a potential system with which to probe kicks associated with PN ejection if we were able to accurately measure the true space velocities of stars A and B. 

\subsubsection{Higher order multiplicity}

Given that we know so little about star B, it is possible that is was initially (or may still be) a binary star as well, thus making the Sh 2-71 progenitor system a quadruple of 2+2 hierarchy. This setup would not significantly affect our scenario except for changing the main-sequence lifetime of B and potentially making the Lidov-Kozai cycles more efficient \citep[e.g.][]{pejcha13,fang18}.

\section{Discussion and conclusions}
\label{sec:discussion}

Based on the inferred properties of stars A and B and a few simplifying assumptions, we have shown that it is plausible that at some point binary A formed a triple system with star B.  The interactions within such a triple system as star B (the nebular progenitor in this scenario) lost its AGB envelope could feasibly have led to the formation of the unusual precessing disc found by \citet{mocnik15} in binary A.  Similarly, mass transfer between star A and binary B could have led to the formation of jets (perhaps blown from the disc in binary A) which directly impacted upon the shaping of the nebula - leading to the ``messy'' morphology of the PN \citep[considered to be a tell-tale sign of triple interactions;][]{akashi17,bear17} as well as the observed shocks \citep{bohigas01}.  Furthermore, the newly identified extended emission regions, lying several arcminutes away from the centre of the PN but approximately aligned with the current positions of binary A and star B, may well be signposts of interactions within binary A or between binary A and the mass lost from star B during the PN formation episode.  

While the proposed triple scenario for Sh~2-71 is apparently plausible, and perhaps even favourable in explaining the nebular morphology, the ultimate test of this hypothesis will likely be improved parallax and proper motion measurements later in the \textit{Gaia} mission. Such measurements will hopefully prove conclusive in assessing the association of star B not only to binary A but to the PN itself. Tracing the PN expansion relative to A and B, for example with two well-separated epochs of high-resolution images, could further elucidate the kinematics and past evolution of the system. Alternatively, one may consider a spectral analysis of star B to check whether it is consistent with a high mass remnant (as implied by the relative evolutionary timescales discussed in Sec.~\ref{sec:triple}) and/or whether its radial velocity is coincident with that of the systemic velocity of the nebula. Understanding the peculiar variability of A could shed more light on its relation with the PN ejection. Similarly, UV observations of binary A could be used to search for the presence of the proposed hot subdwarf nebular progenitor.  Unfortunately, none of these observations would be easy given the dearth of UV observatories and the very faint nature of star B.  However, they could prove critical in understanding the origins of this fascinating nebula.

%While the triple hypothesis for Sh~2-71 is far from being proved, t
In conclusion, the fact that the currently available observations of Sh~2-71 stand-up to the (somewhat circumstantial) tests to which they have been subjected here means Sh~2-71 remains one of the most promising candidates to host (or rather have hosted) a triple central star.

\section*{Acknowledgements}
The authors would like to thank the anonymous referee for their constructive comments and suggestions.  DJ acknowledges support from the State Research Agency (AEI) of the Spanish Ministry of Science, Innovation and Universities (MCIU) and the European Regional Development Fund (FEDER) under grant AYA2017-83383-P.  DJ also acknowledges support under grant P/308614 financed by funds transferred from the Spanish Ministry of Science, Innovation and Universities, charged to the General State Budgets and with funds transferred from the General Budgets of the Autonomous Community of the Canary Islands by the Ministry of Economy, Industry, Trade and Knowledge. OP was supported by Horizon 2020 ERC Starting Grant ``Cat-In-hAT'' (grant agreement \#803158) and INTER-EXCELLENCE grant LTAUSA18093 from the Czech Ministry of Education, Youth, and Sports. This research was also supported by the  Erasmus+  programme  of the European Union under grant number 2017-1-CZ01-KA203-035562.

Based on observations made with the Isaac Newton Telescope operated on the island of La Palma by the Isaac Newton Group of Telescopes in the Spanish Observatorio del Roque de los Muchachos of the Instituto de Astrof\'isica de Canarias.  Based on observations obtained at the Gemini Observatory acquired through the Gemini Observatory Archive, which is operated by the Association of Universities for Research in Astronomy, Inc., under a cooperative agreement with the NSF on behalf of the Gemini partnership: the National Science Foundation (United States), National Research Council (Canada), CONICYT (Chile), Ministerio de Ciencia, Tecnolog\'{i}a e Innovaci\'{o}n Productiva (Argentina), Minist\'{e}rio da Ci\^{e}ncia, Tecnologia e Inova\c{c}\~{a}o (Brazil), and Korea Astronomy and Space Science Institute (Republic of Korea).  Based on data products from observations made with ESO Telescopes at the La Silla Paranal Observatory under public survey programme ID, 177.D-3023.  This research made use of Astropy, a community-developed core Python package for Astronomy \citep{astropy}. Simulations in this paper made use of the REBOUND code which can be downloaded freely at http://github.com/hannorein/rebound.

%%%%%%%%%%%%%%%%%%%%%%%%%%%%%%%%%%%%%%%%%%%%%%%%%%



\bibliographystyle{mnras}
\bibliography{literature}

%%%%%%%%%%%%%%%%%%%%%%%%%%%%%%%%%%%%%%%%%%%%%%%%%%

%%%%%%%%%%%%%%%%% APPENDICES %%%%%%%%%%%%%%%%%%%%%

%%%%%%%%%%%%%%%%%%%%%%%%%%%%%%%%%%%%%%%%%%%%%%%%%%

% Don't change these lines
\bsp	% typesetting comment
\label{lastpage}
\end{document}